\documentclass[conference]{IEEEtran}
\usepackage{cite}
\usepackage{amsmath,amssymb,amsfonts}
\usepackage{algorithmic}
\usepackage{graphicx}
\usepackage{textcomp}
\usepackage{xcolor}

\usepackage{balance}

\usepackage{bm}

\usepackage{enumitem}
\usepackage{hyperref}
\usepackage{url}
\usepackage{environ,lipsum,xcolor}

\usepackage{graphicx}
\ifCLASSOPTIONcompsoc
    \usepackage[caption=false, font=normalsize, labelfont=sf, textfont=sf]{subfig}
\else
\usepackage[caption=false, font=footnotesize]{subfig}
\fi

\usepackage{datatool, filecontents}
\DTLsetseparator{ = }

\def\BibTeX{{\rm B\kern-.05em{\sc i\kern-.025em b}\kern-.08em
    T\kern-.1667em\lower.7ex\hbox{E}\kern-.125emX}}

\begin{document}
\bstctlcite{IEEEexample:BSTcontrol}
\DTLloaddb[noheader, keys={thekey,thevalue}]{mydata}{mydata.dat}

\newcommand{\insertdata}[1]{\DTLfetch{mydata}{thekey}{#1}{thevalue}}

\title{IdeoTrace: A Framework for Ideology \\
Tracing with a Case Study on\\
the 2016 U.S. Presidential Election
}

\author{\IEEEauthorblockN{Indu Manickam}
\IEEEauthorblockA{\textit{Rice University} \\
indu.manickam@rice.edu}
\and
\IEEEauthorblockN{Andrew S.\ Lan}
\IEEEauthorblockA{\textit{University of Massachusetts Amherst} \\
andrewlan@cs.umass.edu}
\and
\IEEEauthorblockN{Gautam Dasarathy}
\IEEEauthorblockA{\textit{Arizona State University} \\
gautamd@asu.edu}
\and
\IEEEauthorblockN{Richard G.\ Baraniuk}
\IEEEauthorblockA{\textit{Rice University} \\
richb@rice.edu}
}

\maketitle
\thispagestyle{plain}
\pagestyle{plain}

\begin{abstract}
The 2016 United States presidential election has been characterized as a period of extreme divisiveness that was exacerbated on social media by the influence of fake news, trolls, and social bots. However, the extent to which the public became more polarized in response to these influences over the course of the election is not well understood. In this paper we propose IdeoTrace, a framework for (i) jointly estimating the ideology of social media users and news websites and (ii) tracing changes in user ideology over time.  We apply this framework to the last two months of the election period for a group of \insertdata{num_total_users} Twitter users and demonstrate that both liberal and conservative users became more polarized over time. 
\end{abstract}

\section{Introduction}
Contrary to President Obama's famous declaration that, ``There is not a liberal America and a conservative America---there is the United States of America", growing evidence points to the large divisions between liberal and conservative Americans. In addition to a widening gap on political issues \cite{PEW2017}, liberal and conservative Americans have become more segregated in terms of geographic location \cite{Martin2018}, and cultural and lifestyle preferences \cite{DellaPosta2015}. There have even been differences found in brain structure \cite{Kanai2011} and basic biological responses \cite{Ahn2014} between liberals and conservatives.

While social media reflects the current divisiveness through the homophily exhibited in online social networks, \cite{Arvidsson2014, Boutyline2017}, there may also be factors unique to social media which are actually exacerbating polarization. For example, a recent study found that the percentage of opinion pieces versus descriptive news in users' news consumption is larger when accessing articles through social media instead of directly visiting news websites \cite{Flaxman2016}. This shift in media diet may be driving social media users further to the left or right as opinion pieces have been shown to increase readers' political bias \cite{Coppock2018}. In addition, social media has become a platform and an amplifier for malicious online actors, including bots and trolls, who are intent on increasing polarization and manipulating users. Several studies have demonstrated the success of bots in moving users with initially moderate views to more extremist positions  \cite{gorodnichenko2018, Ferrara2016, aiello2012, Abokhodair2015, Luceri2019}. It has also been shown that trolls have embedded themselves within social networks on both sides of controversial issues with the intent of amplifying conflict, including in the $\#\text{BlackLivesMatter movement}$ \cite{Stewart2018} and the vaccination debate \cite{Broniatowski2018}.  

The effect of social media on public opinion came to international attention during the 2016 United States presidential election, when it was determined that Russian trolls operating through the organization called the Internet Research Agency (IRA) posed as American voters on online platforms including Facebook, Instagram, and Twitter, with the intent of sowing divisiveness in the election \cite{justice_2018}. Recent studies have examined troubling aspects of social media during the 2016 election, including the activity and social connectivity of Russian trolls \cite{badawy2018b, badawy2018a} and the propagation of fake news \cite{bovet2019,  Grinberg2019}. However, to date there has been limited analysis on whether the polarization between liberals and conservatives actually increased over the election period, potentially in response to these malicious attacks. Given that much of the focus of Russian trolls was on propagating fake news and directing users towards extremist websites, we are particularly interested in measuring polarization based on news media consumption in order to understand whether liberal and conservative users were becoming more driven apart into information filter bubbles. 

The focus of this work is therefore on the development of a framework to jointly estimate the ideology of social media users and the news sources they interact with online, and to trace the evolution of user ideology over time.  Using this framework, we analyze Twitter activity over the final months of the 2016 election in order to detect trends in the polarization of users over time and trace the shift in individual social media users.

 



\subsection{Contributions}
In this paper, we propose IdeoTrace, a framework to jointly estimate the political ideology of news websites and social media users, and to trace the ideology of social media users over time. This approach uses a matrix factorization method to model political ideology without requiring labelled data, allowing this work to be extended to large-scale datasets. IdeoTrace also leverages the  network between users to improve model estimates by imposing that socially connected users share similar ideology. 

After applying the IdeoTrace framework to Twitter posts (tweets) related to the election that were published between September 1 and Election Day, November 8, 2016, we demonstrate that the liberal and conservative clusters of users  became more polarized over time. We do so by showing that the average distance between the liberal and conservative clusters increased over time.

In this paper, the term \emph{users} will be used to refer to social media users, and the term \emph{websites} will be used to refer specifically to websites generating written content on political topics. The terms political \emph{ideology} and \emph{bias} will also be used interchangeably to refer to a given user's orientation towards political issues. We will use standard terminology from U.S.\ politics and refer to the two political ideology groups as \emph{liberals}, or the \emph{left}, and \emph{conservatives}, or the \emph{right}.

\section{Background and Related Work}\label{sec:background}
Previous work in social analysis has either focused on modelling opinion dynamics in a social network over time or on estimating the ideology of news sources and social media users at a single time point.  We detail prior work on both of these problems that we drew upon in formulating the IdeoTrace model. 

\subsection{Ideology Estimation} \label{sec:ideology_estimation}
There is a large body of work in social media analysis that ignores the temporal aspects of ideology, and instead uses statistical models to infer the political ideology of social media users and/or news websites based on features such as users' follow networks and the language in social media posts and news articles. We refer to this body of work as ideology estimation.

Recent works on estimating the political ideology of social media users include \cite{iyyer2018} and \cite{Maynard2012} which infer users' ideology based on the text in their social media posts using an recurrent neural network framework and natural language processing techniques respectively. In \cite{barbera2015},  political ideology is inferred by jointly estimating the ideology of Twitter users and the politicians that they follow using a Bayes ideal-point estimation model. 

Other studies have focused on inferring the political ideology of news websites.  This includes \cite{baly2018}, where the authors designed a support vector machine to classify the political ideology and factuality of websites using input features such as article popularity, web traffic, sentiment scores, and linguistic expressions of the article text.  Another popular approach to this problem is the use of matrix factorization. In \cite{Niculae2015}, the authors create a bipartite graph representing the sections from the entire corpus of President Obama's official speeches which were quoted by various news sources.  The authors used a matrix completion algorithm to learn the latent features capturing which news sources are likely to quote certain sections of President Obama's speeches. The first two features of this latent space were found to roughly correspond to political ideology (liberal/conservative) and the type of news source (mainstream/independent). 

More recent papers have also focused on jointly estimating  user ideology and news source ideology using matrix factorization. In \cite{Shu2019}, the authors developed a shared matrix factorization model that trained on a dataset with article text labeled as factual or fake news in order to jointly estimate the latent features of social media users, article text, and news sources. The model is used to detect which news sources are more likely to produce fake news and estimate users' individual susceptibility to spreading fake news. Using an approach most similar to that detailed in this paper for ideology estimation, Lahoti et al.\cite{Lahoti2018} used a matrix factorization model to jointly estimate user and news source ideology for Twitter users, with an additional penalty enforcing smoothness over the retweet network. IdeoTrace is an extension of this work that, in addition to jointly estimating website and user ideology using unlabeled data, also considers user ideology at each point on time.

\subsection{Opinion Dynamics Models} \label{sec:opinion_dynamics}
There is also a large body of work around opinion dynamics, which aims to trace the evolution of users' opinions over time. Note that while this paper is primarily concerned with the estimation of users' political ideology, the term \emph{opinion} is a generalization of the concept of political ideology, and is defined as an individual's ``cognitive orientation towards an object", such as an event, topic, or another individual, and can be represented using a real-valued scalar or vector \cite{Proskurnikov2017}.

Opinion dynamics models typically either ignore influences on ideology external to the social network between users, or treat these influences as known inputs. The early, well-established work on this problem centered around the development of theoretical models that, given a social network of users with an initial distribution of opinions, use rule-based updates often inspired by physical or biological networks to estimate user opinion over multiple time steps. The models are then shown in simulation to converge to particular distributions based on the network and update rules. For example, in \cite{vicario2017} it was shown that modelling confirmation bias by moving users with similar opinions closer together each simulation step and breaking the network link between them otherwise, often results in the formation of a bimodal opinion distribution. For a review of popular opinion dynamics models, we refer to \cite{Proskurnikov2017}. 


While the theoretical work in opinion dynamics has been important for understanding the type of update rules and network structures which lead to consensus or bimodal opinion distribution, the emergence of social media has shifted the focus of recent papers from theoretical work to using observational data to trace user ideology. 

In one of the first works to trace long-term political ideology, Garimella et al. \cite{Garimella2017a} measured the polarization of Twitter users over an eight year period. The authors labeled the accounts of prominent politicians as liberal or conservative, and users were obtained for the study by selecting among the followers of these politicians. Polarity was measured by modelling the likelihood at each point in time of a user retweeting liberal versus conservative accounts.  In a separate experiment, they also measured polarization based on the differences in hashtag usage between liberal and conservative users.  Through both measures they authors demonstrated an increase in polarization over the eight year period by 10-20\% \cite{Garimella2017a}. 

A recent paper, \cite{bovet2018}, also examined temporal trends over the 2016 election period, and showed that the activity levels of Twitter users provides a reasonable estimate of candidate favorability in opinion polls. The authors determined that the co-occurrence network of hashtags included in election-related tweets was composed of two clusters, one consisting of hashtags that were pro-Clinton or anti-Trump, and the other of hashtags that were pro-Trump or anti-Clinton. The paper used a classifier to categorize users as pro-Clinton or pro-Trump using these labeled hashtags, and demonstrated that the percentage of users favoring either candidate closely tracked with \emph{New York Times} opinion polls for Clinton and Trump respectively over the five month period leading to the election. 


\section{Approach}\label{sec:approach}
In this section, we develop IdeoTrace, a model that uses matrix factorization techniques to jointly estimate (i) the ideology of websites (constant over time) and (ii) the evolution of user ideology over time based on users' social media activity.

\subsection{Assumptions} \label{sec:assumptions}
 We first detail the assumptions governing the online behavior of social media users that we utilize in the design of IdeoTrace.
 
\begin{enumerate}[label=(A\arabic*),ref=A\arabic*]
    \item \textit{Each website has an overall bias or ideology that affects the viewpoints expressed in its published content, and similarly each user has an underlying bias or ideology that affects her online behavior.} We can view a website's ideology as a summation of the political ideology expressed over all articles produced by the organization. The existence of an underlying news website ideology that affects the organization's overall behavior is supported by studies which were able to determine news source ideology based on factors such as selection bias and linguistic features of article text \cite{Shu2019, Niculae2015, Gentzkow2010}. The existence of user bias is also well supported by studies finding evidence of confirmation bias in users' news consumption via social media \cite{Bakshy2015}. This assumption is clearly necessary for our approach of modelling political ideology as the latent variable affecting user behavior on social media.  \label{A1}
    
    
    \item \textit{Users generally share articles that reflect their political viewpoint.} While the dataset used in this experiment includes cases where users linked an article along with a comment in the post indicating they found the viewpoints in the article to be absurd, in general these instances appeared to be less common. This assumption is further supported by studies such as \cite{Garimella2018} which found a strong correlation between the average ideology expressed in Twitter users' media consumption to the average ideology of articles shared by users. This assumption allows us to model the likelihood of a website being shared by a particular user as an inner product between the representation of the user's ideology and the website's ideology.  \label{A2}
    
    \item \textit{Users form social networks with other users who reflect their viewpoint.} This is an example of homophily, the tendency of individuals to connect with others who hold similar opinions to their own; its presence in social networks has been well documented  \cite{McPherson2001, Arvidsson2014, Halbertstam2016}. Based on this assumption we can enforce in our model that the estimated ideologies for users should be smooth over their social network. \label{A3}

    
    \item \textit{The ideology of  news websites does not vary over the election period.} Although there are studies that have found evidence of a drift in media ideology, e.g. \cite{Gasper2011}, we can expect that a media organization, which is comprised of multiple journalists, will be slower to shift in overall ideology in comparison to a single user. This assumption allows us to treat the website ideology as fixed in the IdeoTrace framework and track changes in user ideology over time. \label{A5}

\end{enumerate}

\subsection{Model of Social Media Behavior} \label{sec:model}
We consider a set of $N$ social media users posting tweets with links to articles produced by a set of $M$ news websites (e.g. nytimes.com, cnn.com, breitbart.com). Using Assumption \ref{A1} we represent the political ideology of both users and websites as vectors that lie in a $K$-dimensional space.  Based on Assumption \ref{A2}, it follows that if the vector representation of a given user's political ideology is closely aligned with the vector representation of a website's ideology, the user is more likely to share an article produced by the website. The proposed statistical model therefore treats the probability that a user shares an article on social media at a particular time as a function of (i) the inner product between the current ideology of the user and the ideology of the news source, (ii) the popularity of the news source, and (iii) the user's activity level on social media. 

We formulate the problem as follows. We are given a binary matrix $Y \in \{0,1\}^{M \times N}$ where entry $y_{ij} = 1$ indicates that user $j$ shared an article produced by website $i$, and $y_{ij} = 0$ otherwise. Let the matrix $C \in \mathcal{R}^{N \times K}$ be composed of row vectors $\bm{c_j} \in \mathcal{R}^K$ that represent each user $j$'s political ideology.  Similarly, we let the matrix $W \in \mathcal{R}^{M \times K}$, which is composed of row vectors $\bm{w_i} \in \mathcal{R}^K$, denote the political ideology of each website $i$. The vector $\bm{\mu} \in \mathcal{R}^M$ captures the overall popularity of every website, and the vector $\bm{\nu} \in \mathcal{R}^N$  captures the overall activity level of each user.  Treating each entry $y_{ij}$ as a Bernoulli random variable, we can then write the probability that user $j$ shares an article produced by website $i$ as follows:
%
\begin{align}
    z_{i,j} &= \Phi({\bm{w_i} \bm{c_j}^T + \mu_i + \nu_j}) \quad \forall i,j \label{eq:z} \\ 
    y_{i,j} &\sim \text{Ber}(z_{i,j}) \label{eq:y}
\end{align}
%
In \eqref{eq:z}, $\Phi(\cdot)$ is an inverse link function that maps the inner product between $\bm{w_i}$ and $\bm{c_j}$, which represents the alignment in ideology between the user $j$ and the website $i$, to the success probability of a Bernoulli random variable $y_{i,j} \in \{0,1\}$. For ease of computation, the link function used is the inverse logit function, which is defined as $\Phi(x) = \frac{1}{1 + \exp(-x)}$. Note that in this expression $\mu_i$ is a single entry from $\bm{\mu}$ which captures the bias of a single website $i$, and similarly $\nu_j$ is a single entry from $\bm{\nu}$ capturing the bias of a single user. 

As a result of elements of W and C being able to take on both positive and negative values, if a particular set of vectors $\bm{c_j}$ and $\bm{w_i}$ have the same sign in each dimension, the user is more likely to share an article from that website. This results in the sign of each dimension corresponding to two sides of the political ideology spectrum. 

\subsection{Tracing User Ideology} \label{sec:tracing}
We now extend the model described in Section \ref{sec:model} to consider the case where user ideology changes over time. Let $T$ denote the total number of time steps. In this new framework, at each time point $t$, we are given a binary matrix $Y^t \in \{0,1\}^{M \times N}$ that represents the aggregate set of websites shared by each user since the previous time step. We treat the ideology of websites, $W$ as fixed based on Assumption \ref{A5}. We then model the ideology of all users at time $t$ as the matrix $C^t$. To account for the differences in user activity and website popularity over time, we treat the website bias vector, $\bm{\mu}$, and the user bias vector, $\bm{\nu}$, as time-varying as well.

\subsection{Joint Estimation of User and Website Ideology} \label{sec:estimating}
Given the framework described in Sections \ref{sec:model} and \ref{sec:tracing}, we now describe the methodology used to estimate the set of user ideology matrices $\{C^t\}_{t=0}^T$ and the website ideology $W$ given the set of observation matrices  $\{Y_t\}_{t=0}^T$. Recall that $\{Y_t\}_{t=0}^T$ is the set of binary matrices that represent which websites each user shared at each time point. We use the following loss function that minimizes the negative log likelihood of the observed data:
\begin{align}
    \min_{\{C^t,\mu^t, \nu^t\},W}& \sum_{t=0}^T \Bigg(\sum_{i,j} -\log B_{i,j}^t p\Big(y^t_{i,j}|\bm{w_i},\bm{c_j}^t,\mu_i^t,\nu_i^t\Big)\Bigg) \nonumber\\
    &+ \frac{\gamma}{2}||W||^2_F + \frac{\gamma}{2} \sum_{t=0}^T ||C||^2_F \nonumber\\ 
    &+ \sum_{t=0}^T \Bigg( \lambda \text{tr}\Big((C^t)^T \mathcal{L_R} C^t\Big) +  \tau ||C^t - C^{t-1}||^2_F \Bigg) \label{eq:loss}.
\end{align}
In this expression, $p(y^t_{i,j}|\bm{w_i},\bm{c_j}^t,\mu_i^t,\nu_i^t)$ is derived based on \eqref{eq:z} and \eqref{eq:y}, and can be written as $\Phi(2(y_{i,j}^t - 1)(\bm{w_i}(\bm{c_j})^T + \mu_i^t + \nu_j^t))$. The matrix $B^t$ is a weighting function which is used to assign a larger cost in the loss function to the set of websites that users shared (entries where $y^t_{i,j} = 1$) over the set of websites users did not share (entries where $y^t_{i,j} = 0$). The term \emph{tr} refers to the trace of the matrix and  $\mathcal{L}_R$ represents the Laplacian of the user social network and is used to enforce the assumption of homophily over the social network (see Assumption~\ref{A3}); this is expanded upon below.

The matrix $B^t \in \mathcal{R}^{M \times N}$ is used to differentiate the loss function between positive and negative labels. Social media datasets are an example of implicit data, where only positive labeled entries are observed, and we are unable to differentiate between negative and unlabeled data. In other words, the fact that a user did not share an article from a particular website could either indicate that the user disagreed with the content of the article, or that the user agrees with the website ideology but was unaware of the website or did not come across articles they wished to share produced by the website. Following the approach outlined in \cite{Yu2017} and \cite{Hsieh2015}, we adjust for the higher uncertainty associated with negative labels by setting $B^t_{i,j}=1$ for entries where $y^t_{i,j}=0$, and setting $B^t_{i,j}=\beta$ for entries where $y^t_{i,j}=1$, where $\beta > 1$ is a constant determined through parameter tuning. 

The expression $ \text{tr}( (C^t)^T \mathcal{L_R} C^t)$, which we refer to as the {\em graph penalty}, then enforces the estimate of $C^t$ should be smooth over the social network graph, meaning that if user $j$ is influenced by user $k$, then the estimate for $\bm{c_j^t}$ should be close to the estimate for $\bm{c_k^t}$. We assume that the relationship between users can be captured by a social network which we represent using an undirected, unweighted graph, where users correspond to nodes in the graph and an edge between user $j$ and user $k$ indicates that at least one of the users is influenced by the behavior of the other. The Laplacian matrix $\mathcal{L}_R \in \mathcal{R}^{N \times N}$ is defined as the difference between the degree matrix, a diagonal matrix where each entry $D_{j,j}$ is the edge degree of user $j$, and the adjacency matrix, which is a binary matrix where entry $A_{j,k} = 1$ if an edge exists between  users $j$ and $k$ and $A_{j,k} = 0$ otherwise. The graph penalty, also referred to as the Laplacian quadratic form, is known to be equal to the sum of differences in $\bm{c_j}$ and $\bm{c_k}$ between all pairs of users $j$ and $k$ that are connected in the graph \cite{Spielman2017}. 

As is normal in ridge regression, the $L_2$ penalties, $||C^t||^2_F$ and $||W||^2_F$, are used to induce interpretability in the results by constraining the magnitude of the estimates of $\{C^t\}$ and $W$. Finally to ensure smoothness over adjacent time points we also impose a loss penalty using the squared Frobenius norm on the difference in magnitude between adjacent $C^t$ matrices. 


Given that the loss function \eqref{eq:loss} is nonconvex, the estimates for the parameters were determined by finding a local minima. Point estimates for $W$, $\{C_t\}$, $\{\mu_t\}$, and $\{\nu_t\}$ were determined using the Adam stochastic gradient descent method as implemented in TensorFlow \cite{tensorflow2015}. The values for $\beta$ and the regularization parameters $\gamma$, $\tau$ and $\lambda$ were determined using cross validation. 


\section{Experiments} \label{sec:experiments}
We now demonstrate the performance of IdeoTrace on a real-world social media dataset. Given that the data is primarily clustered into a set of liberal users and websites and a set of conservative users and websites, we found in practice that setting the latent ideology dimension to $K=2$ yielded results that were most interpretable. 

We ran three sets of experiments:
\begin{itemize}
    \item In Section \ref{sec:eval_sites}, we compare the model estimates against the set of ground truth labels for the ideology of users and websites.
    
    \item In Section \ref{sec:validation}, we measure the performance of the model in predicting the posting behavior of a new set of Twitter users using the estimates for $W$ and $\{\mu_t\}$ computed from training.
    
    \item In Section \ref{sec:tracing_results}, we determine whether users became more polarized over time based on the model estimates of user ideology.
\end{itemize}

\subsection{Dataset and Processing} \label{sec:dataset}

The dataset used for this experiment is a publicly available collection of tweets that were posted between July 13, 2016 and November 8, 2016 \cite{littman2016}.  The dataset was originally gathered  by Littman et al.\ using Twitter's Streaming API to filter out tweets that included terms related to the election such as ``election2016", ``election", ``clinton", ``kaine", ``trump", and ``pence" in the Tweet text, as well as terms related to specific election events such as the three debates. In compliance with Twitter's developer policy, the dataset contains only the ID numbers for the Tweets. The complete Twitter dataset was collected through Twitter's Rest API using a software package called Hydrator\footnote{\url{https://github.com/DocNow/hydrator}}. Due to Twitter's restrictions, any deleted messages or messages posted by accounts that were deleted prior to the date the data was downloaded, which began in September 2018, were not included in the dataset.  As a result the dataset likely does not include tweets from accounts that were flagged as being Russian trolls and many bot accounts which were deleted immediately after the election.

To reduce the set of websites to English-based news sources, only tweets that were written in English were included. From this reduced set of tweets, we constructed the set of observation matrices $\{Y^t\}_{t=0}^T$ and the Laplacian of the retweet network, $\mathcal{L}_R$. To ensure that we have have at least one observation per time point, we defined a time point as a two week period from September to November of 2016, and select a set of \insertdata{num_total_users} users who shared at least four articles in each time period from a set of \insertdata{num_websites} websites. The full set of websites posted by these this particular set of users is larger, but the set was narrowed to the most popular websites to ensure that there was a sufficient number of observations per website. 

To construct $\{Y^t\}_{t=0}^T$, we consider the set of tweets containing a URL link. Note that because a large portion of the URLs were shortened links that redirect through link shortening service such as bit.ly or ow.ly, the Python requests library was used to extract the original domain name.  To ensure that the URL link was to a news article, we filtered out websites that typically link to videos, forum discussions, blog posts, and other unrelated sites such as wordpress.com, youtube.com, reddit.com, wordpress.com, vimeo.com, instagram.com, facebook.com,  and amazon.com.  To construct $Y^t$, we then aggregate the set of URL domains that each user shared links to over the time period $t$ and form the results into a binary matrix.

On Twitter, the social network of users is typically represented by users' follow networks, or by the retweet network. Note that retweet is a term used on Twitter for when a user posts the content produced by another user without altering the original content. We can therefore in general consider a retweet as an indication that the user agrees with the content posted by the other user, and therefore most likely the pair of users have similar political ideologies. Based on the results of \cite{Lahoti2018}, where it was determined in performing joint estimation of user and website ideology that use of the retweet networks resulted in improved performance over use of the follower network, in this paper we also used the retweet network to represent how users influence each other. The retweet graph that is constructed based on this network is an undirected graph where nodes represent users, and where an edge between user $j$ and user $k$ indicates that user $j$ has posted at least one retweet from user $k$ or vice versa over the entire time period.

\subsection{Ground Truth Labels} \label{sec:ground_truth}
We require a set of ground truth labels of user and website ideology to quantify the performance of the ideology estimates produced by IdeoTrace. Given that we do not have additional information on the self-identified bias of websites as well as Twitter users, we rely on other sources including expert labels to determine the ground truth ideology values.

\subsubsection{Website Labels}
For news websites, we use the set of labels on website ideology produced by the organization Media Bias\textbackslash Fact Check \footnote{\url{wwww.mediabiasfactcheck.com}}, which has served as a resource for expert labels on website ideology for related previous work including \cite{badawy2018a, baly2018}. The organization evaluates media sources based on a combination of qualitative and quantitative factors such as biased wording, factuality, story choices, and political affiliation to categorize the website's overall political ideology as extreme right, right, right-center, center, left-center, left, or extreme left. 

\subsubsection{User Labels}
Given that 
Twitter users are anonymous, we do not have access to additional information to determine their political ideology. We instead used the same approach outlined in \cite{Lahoti2018} where we treat the average ideology  of the set of websites shared by each user, as measured by the estimate of $W$, as the ground truth ideology.  To obtain a scalar estimate of the ground truth, we first project $W$ using principle component analysis (PCA) onto a single dimension and then compute the average ideology per user.

\subsection{Evaluation of User and Website Ideology Estimates} \label{sec:eval_sites}
To improve computation speed we split the users into \insertdata{num_folds} sets of approximately \insertdata{num_approx_users_per_group} users and ran the IdeoTrace framework separately on each set. 

We evaluate the accuracy and interpretability of the ideology estimates produced by IdeoTrace by comparing the estimates of $W$ and $\{C^t\}$ for each set of users against the ground truth ideology labels using both visual examination and by computing the correlation coefficient between the model estimates and ground truth. 

\begin{figure}[tbp]
\centerline{\includegraphics[width = \columnwidth]{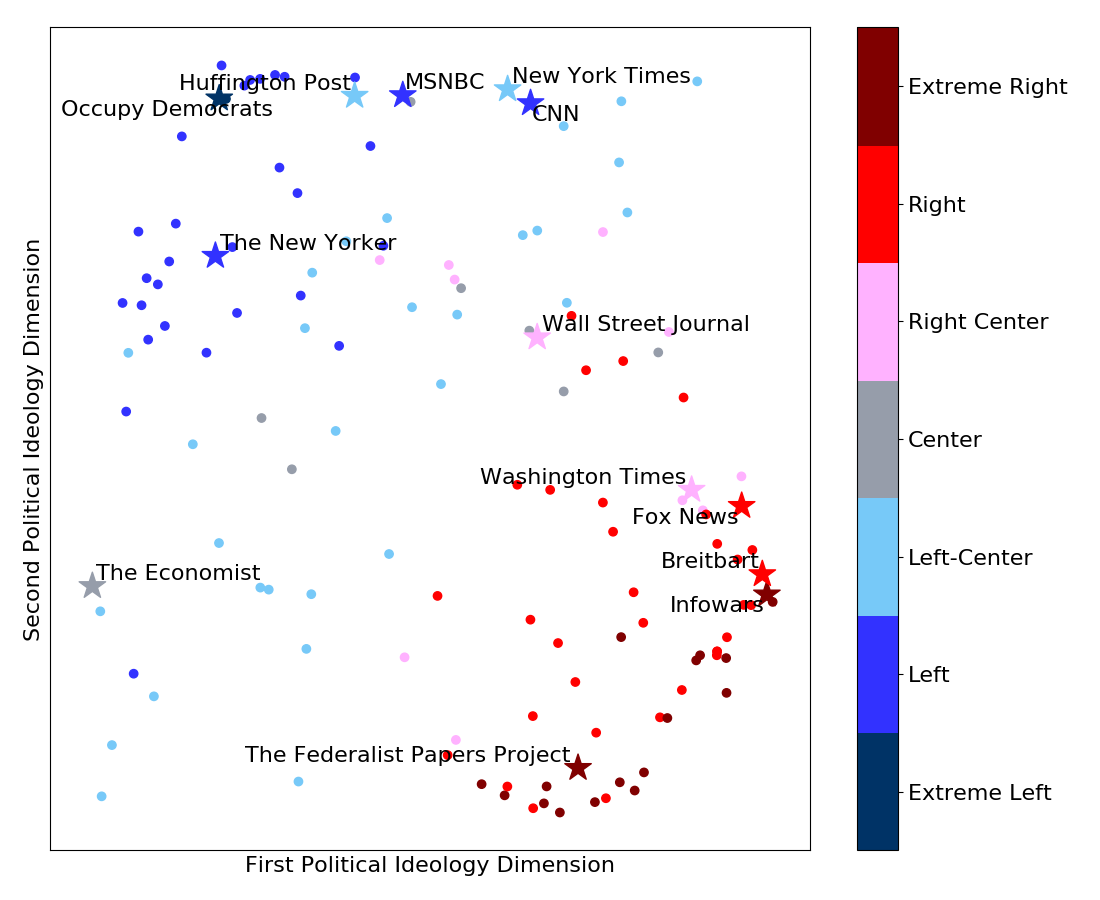}}
\caption{A visualization of the rows of $W$, which represent the estimated political ideology of each website. The axes in the image correspond to the two latent features representing political ideology. Each website is shaded based on the ground truth label.  The figure shows a clear separation between conservative and liberal websites. Examples of well-known news sources such as \emph{The New York Times} and \emph{Fox News} are highlighted with stars.}
\label{fig:W}
\end{figure}

To evaluate the website ideology $W$, we first plot the model estimate for $\{\bm{w_i}\}_{i=1}^M$, the ideology of individual websites, as shown in Figure \ref{fig:W}. The results shown in this image are the IdeoTrace estimate of $W$ after training on a single set of \insertdata{num_approx_users_per_group} users.  The figure displays \insertdata{num_websites_graphed} of the \insertdata{num_websites} websites included in the dataset for which there exist ground truth labels provided by Media Bias/Fact Check. Given that $K = 2$, the ideology of each website exists in a 2-dimensional space so we can directly visualize the results. The coloring of each website indicates the ground truth label.  It is visually clear from the figure that IdeoTrace learns a separation between conservative and liberal websites, and also learns to separate extremely biased websites from slightly biased websites. 


Using the Spearman rank correlation \cite{Zwillinger2000}, we found that the correlation between ground truth and the estimate website ideology, averaging over the \insertdata{num_folds} sets of users, is \insertdata{spearman_corr}. This indicates that the estimated ideologies align fairly well with the ground truth labels.  The Spearman correlation was used in particular as this metric determines the strength of the monotonic relationship between the estimated ideology, which lies on a continuous interval, and the ground truth labels, which is an ordinal set, based on the rank ordering of the data. Values of the coefficient close to 1 indicate a monotonically increasing relationship between ground truth labels and the model estimates, and values close to -1 indicate a monotonically decreasing relationship.

\begin{figure}[tbp]
\vspace{-1cm}
\centerline{\includegraphics[width = \columnwidth]{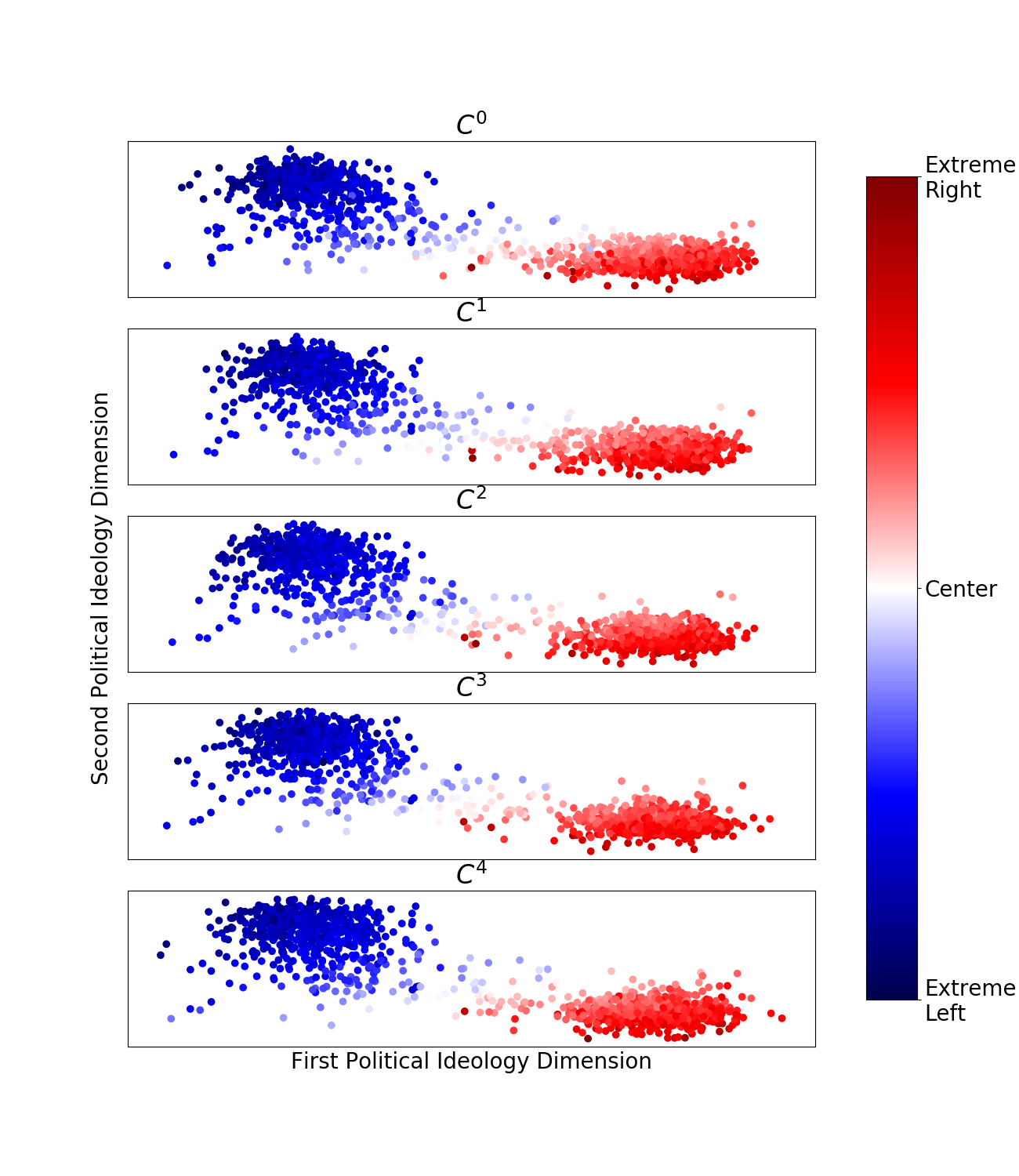}}
\vspace{-1cm}
\caption{A visualization of the columns of $C^t$, which represent the political ideology of each user, at each time step $t$. Each user is shaded based on the ground truth label. 
}
\label{fig:C}
\end{figure}

We also evaluate the accuracy and interpretability of the model's ideology estimate for each user, $\{\bm{c_j^t}\}_{j=1}^N$. In Figure \ref{fig:C}, we visualize the estimated ideology of each user from one user set as represented in the rows of $C^t$ at each time point $t$. Each data point in the figure is colored according to the associated user's ground truth ideology at that point in time.  It is again visually clear from the figure that for all time points the liberal users and the conservative users form two distinct clusters, with few users positioned in between both clusters. 

We also measured the linear relationship between the estimated user ideology value and the ground truth estimate. For this analysis we used the Pearson correlation coefficient, as both the estimated user ideology and ground truth labels lie in continuous intervals. The Pearson correlation coefficient between the estimated user ideology and the ground truth label averaged over all time points and over all sets of users was \insertdata{pearson_corr}, further demonstrating that the estimated values are highly correlated with the ground truth.

This analysis demonstrates that the latent factor in the model corresponds to political bias, and that the estimates produced by IdeoTrace for website and user ideology match ground truth labels. This analysis also confirms that the political bias of the readers of a website serve as a reasonable measurement of the political bias of the website, a finding which has also been supported in a previous study \cite{Ribeiro2018}.


\subsection{Predicting the Ideology of Unobserved Users}\label{sec:validation}
We validate the model against a set of social media users unobserved by the model in training. Using the model estimates produced by training on one of the \insertdata{num_folds} sets of users, we predict user behavior on a second set. This process is repeated for four unique pairs of training and testing sets of users and the results are averaged over all pairs. Note that because IdeoTrace relies in part on the social network to measure ideology, we use the same number of users in both the training and validation set.

For each time step $t_{\textrm{pred}} > 0$ we fix the values of $W$ and $\{\bm{\mu}^t\}_{t=0}^T$ which were determined in training, and run the model on $t < t_{\textrm{pred}}$ to estimate the values of $C_\textrm{val}^{t_{\textrm{pred}}}$ and $\bm{\nu}_\textrm{val}^{t_{\textrm{pred}}}$.  For time $t=0$ we estimate $C_\textrm{val}^{t_{\textrm{pred}}}$ and $\bm{\nu}_\textrm{val}^{t_{\textrm{pred}}}$ by setting the estimate for all users to the average of the training estimates, $C^0$ and $\bm{\nu}^0$. We evaluate prediction performance using the F1 score against two baselines.  

The first baseline is based on the Rasch model \cite{Rasch1993}, which models $p(y_{i,j} = 1) = \Phi(\alpha (\mu_i - \nu_j)$. The Rasch model was formulated based on item response theory and is a popular model for applications such as predicting student performance on academic tests. In the Rasch model, each website $i$ and each user $j$ is represented using a single parameter,   $\mu_i$ and $\nu_j$ respectively. The second baseline is a static version of IdeoTrace which does not use the retweet network and assumes that the user ideology matrix, $C$, is constant over time. 

\begin{table}[tbp]
\begin{center}
\label{tab:f1}
\begin{tabular}{|c|c|}
\hline
\textbf{Model}&\textbf{F1} \\
\cline{1-2} 
$\textbf{IdeoTrace}$ & \insertdata{ideotrace_predict} \\ \hline
Rasch & \insertdata{rasch_predict} \\ \hline
Static MF & \insertdata{static_predict} \\ \hline
\end{tabular}
\end{center}
\caption{Comparison of the F1 mean and standard deviation on predicting the behavior of unobserved users.}
\end{table}

The results are shown in Table \ref{tab:f1}, which shows that IdeoTrace outperforms both baselines.



\subsection{Tracing User Ideology}\label{sec:tracing_results}
We now examine the evolution of user ideology over time for both the liberal and conservative groups of users to determine whether these two groups became more polarized. In particular, we examine whether the centers of the liberal and conservative clusters moved further apart over time. Although the time period over which we are examining user ideology, from September 1 to November 8, 2016, is fairly brief, we were still able to detect trends in user behavior. We ran the analysis separately on each set of users and average the results to determine the overall increase in polarization. 

\begin{figure} [tbp]
    \centering
  \subfloat[Distance between clusters\label{subfiga}]{%
       \includegraphics[width=0.5\columnwidth]{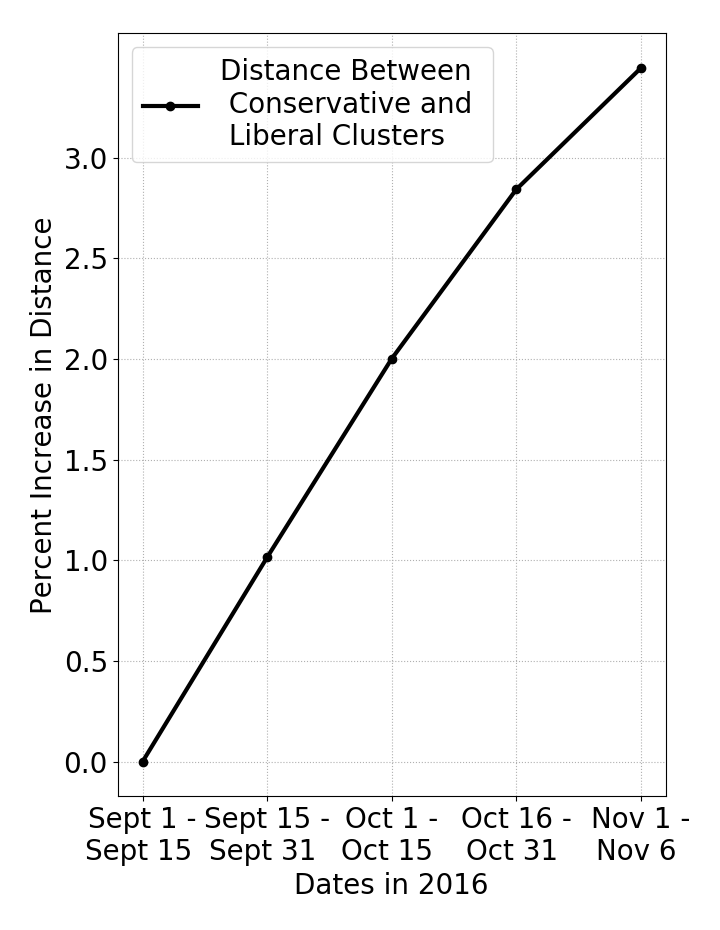}}
    \hfill
  \subfloat[Shift towards extremism\label{subfigb}]{%
        \includegraphics[width=0.5\columnwidth]{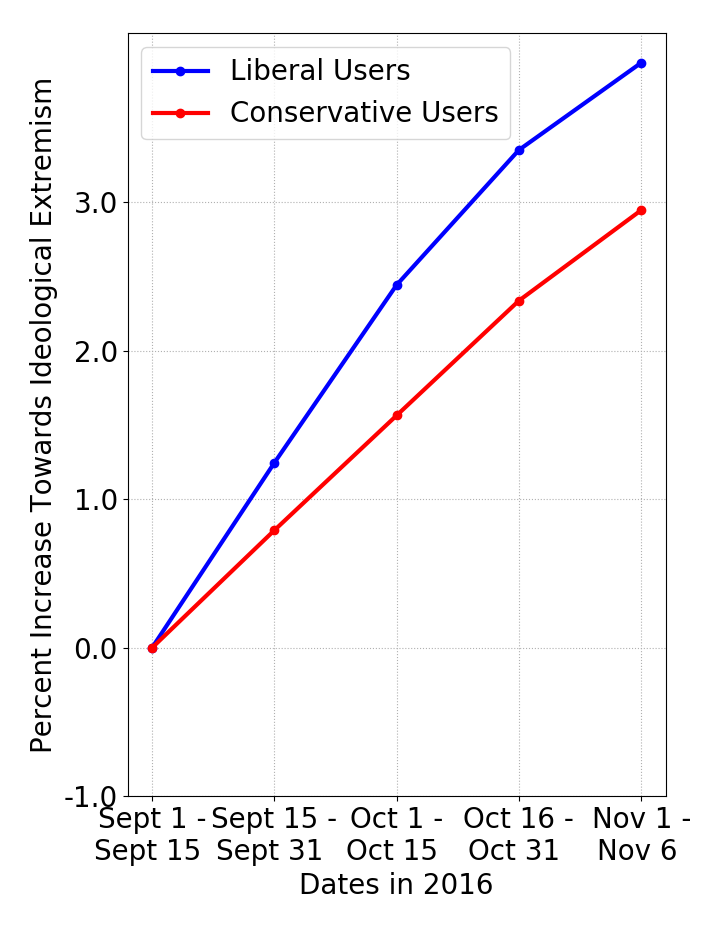}}
  \caption{The (a) percentage increase in distance between liberal and conservatives clusters and the (b) percentage increase towards extremism of liberal and conservative clusters over the course of the election as measured by IdeoTrace.}
  \label{fig:polarization_time} 
\end{figure}

We first cluster the users into the liberal and conservative groups by running a K-means algorithm on the estimates of $C$ at time 0. Using this labeling approach, on average \insertdata{perc_cons_users}\% of each set consists of conservative users, and \insertdata{perc_lib_users}\% consists of liberal users. We then compute the distance between the average value of $C^t$ within the liberal cluster and the average value of $C^t$ within the conservative cluster for each time point $t$. In Figure \ref{subfiga} we plot the percentage increase in the distance between the two cluster means relative to time $t=0$.  From the figure, we can see a steady increase of up to \insertdata{perc_inc_polarization}\% in polarization from September 1 to Election Day. 

We then examine the ideological shifts in the liberal and conservative groups.  Using PCA, we project the estimates of $\{\bm{c_j}\}_{j=1}^N$ onto a single dimension where one direction of the axis is associated with liberalism and the other direction is associated with conservatism. After computing the scalar cluster mean values, we plot the percentage increase towards extremism of both groups as shown in Figure \ref{subfigb}. We can see an overall shift in the liberal cluster towards becoming more liberal by \insertdata{per_shift_lib}\% and an overall shift in the conservative cluster towards becoming more conservative by \insertdata{per_shift_cons}\%. Using the dependent t-test, the p-values for these results were found to be $< 0.001$ for both the liberal and conservative clusters.


\section{Discussion and Future Work}

 In this paper, we presented the IdeoTrace algorithm, which uses matrix factorization to jointly estimate the latent political ideology of both social media users and websites, and also trace the change in political ideology of users over time. We analyzed the performance of the algorithm on a set of Twitter users who shared news articles during the 2016 U.S.\ presidential election.

We demonstrated that the estimates produced by IdeoTrace for news website ideology closely align with the expert-produced ideology labels, and the estimates for Twitter users also appear reasonable based on the set of websites each user shared.  Our observation that Twitter users formed tightly clustered news media bubbles, where users primarily shared articles from either the space of conservative news outlets or the space of liberal news outlets, supports the results of previous papers including the work by Lahoti et al.\ \cite{Lahoti2018}.  A second key finding from this paper is that the liberal and conservative clusters became more polarized over time by moving further apart in the ideological space. This claim is also supported by the prior study by Garimella et al.\ \cite{Garimella2017a}, which also found, using a different metric and approach from that detailed in this paper, an increase in polarization between liberal and conservative users over the 2016 election. 

Increasing polarization should be an issue of public concern.  Recent studies have demonstrated correlation or even direct causation between Twitter activity and the rise of extremist and violent movements.  Detection of polarized behavior on social media directly preceded violent protests in numerous societies including Baltimore \cite{Korolov2016}, Egypt \cite{Weber2013}, and Venezuela \cite{Morales2015}. More troubling, there have been multiple events where fake news posts directly triggered mob rage and violence against minority groups in countries such as Bangladesh \cite{Naher2018} and Myanmar \cite{Mozur2018}.

In response to the growing threats of polarization, extremism, and fake news, researchers are proposing new methods of intervention by, for example, increasing exposure to diverse opinions on social media \cite{Tucker2018, Garimella2017b} and using bots to intervene when racist language is detected  \cite{Munger2016}. Under pressure from governments and social media users, platforms such as Facebook and Twitter are also proposing changes to their system such as adding content moderators and automated algorithms to flag suspicious accounts and content. However, it is unclear at present whether these interventions are actually effective, or whether these interventions are effectively targeting individuals who are most at risk for becoming more extremist. Therefore one important potential application of the IdeoTrace framework could be to serve as a scalable tool for tracing changes in user ideology and enable researchers to quantitatively evaluate the success of these large-scale interventions in combating the negative effects of social media.

\section*{Acknowledgments}

This work was supported in part by ONR grant 
N00014-17-1-2551.

\bibliographystyle{IEEEtran}
\balance
\bibliography{references}

\end{document}